\documentstyle{lamuphys}
\makeatletter
\let\chapter\hid@chapter
\makeatother
\def\be{\begin{equation}}
\def\ee{\end{equation}}
\def\bea{\begin{eqnarray}}
\def\eea{\end{eqnarray}}
\def\ep{\epsilon}
\def\l{\cal L}
\begin{document}
\pagenumbering{arabic}
\title{Matrix Membranes and Integrability}   

\author{Cosmas\,Zachos\inst{1}, David\,Fairlie\inst{2}, and
Thomas\,Curtright\inst{3} }

\institute{High Energy Physics Division, Argonne National Laboratory, 
Argonne,\\  
IL 60439-4815, USA \qquad{\sl zachos@hep.anl.gov}      
\and
Department of Mathematical Sciences, University of Durham, 
Durham,\\  DH1 3LE, UK  \qquad{\sl David.Fairlie@durham.ac.uk}
\and 
Department of Physics, University of Miami, Box 248046, Coral 
Gables,\\  FL 33124, USA \qquad{\sl curtright@phyvax.ir.Miami.edu}  }
\maketitle

\begin{abstract}
This is a pedagogical digest of results reported in [\cite{cfz}],
and an explicit implementation of Euler's construction for the solution of
the Poisson Bracket dual Nahm equation. But it does not cover 9 and 
10-dimensional systems, and subsequent progress on them [\cite{latestdbf}].
Cubic interactions are considered in 3 and 7 space dimensions, respectively,
for bosonic membranes in Poisson Bracket form. Their symmetries and vacuum
configurations are explored. Their associated first order equations are
transformed to Nahm's equations, and are hence seen to be integrable, for the
3-dimensional case, by virtue of the explicit Lax pair provided. Most 
constructions introduced also apply to matrix commutator or Moyal Bracket 
analogs. 
\end {abstract}

\section{\bf Introduction}

A proposal for non-perturbative formulation of M-theory [\cite{banks}] has 
encouraged a reappraisal of matrix membrane theory 
[\cite{collins}, \cite{hoppe1}]. Symmetry features of 
membranes and their connection to matrix models 
[\cite{hoppe1}, \cite{floratos3}, \cite{ffz}, \cite{floratos2}, 
\cite{ffzjmp}] have been appreciated for quite some time. 
Effectively, infinite-$N$ quantum mechanics matrix models 
(presented as a restriction of SU$(\infty)$ Yang-Mills theories) 
amount to membranes, by virtue of the connection between SU($N$)
and area-preserving diffeomorphisms ({\em SDiff}) generated by Poisson 
Brackets: in these, ``color" algebra indices Fourier-transform to ``membrane" 
sheet coordinates. The two are underlain and linked by {\em Moyal Brackets},
the universal associative generalization of Poisson Brackets.

Below, we care to introduce novel Poisson Bracket interactions for a bosonic 
membrane embedded in 3-space
\be
{\l}_{IPB} = \frac{1}{3} \ep^{\mu\nu\kappa} X^\mu \{ X^\nu ,\ X^\kappa \}~,
 \label{termPB}
\ee
which are restrictions of the Moyal Bracket generalization
\be
{\l}_{IMB} = \frac{1}{3} \ep^{\mu\nu\kappa}X^\mu \{ \{ X^\nu ,\ X^\kappa \}\}~,
 \label{termM}
\ee
which, in turn, also encompasses the plain matrix commutator term
 \be
 {\l}_{IC} = \frac{1}{3} \ep^{\mu\nu\kappa} X^\mu [ X^\nu ,\ X^\kappa ]~.
 \label{termC}
 \ee

The structure of (\ref{termPB}) may be recognized as that of the interaction 
term   
$\ep_{ijl} \phi^i \ep^{\mu\nu}  \partial_\mu \phi^j \partial_\nu \phi^j $ 
of the 2-dimensional SO(3) pseudodual chiral $\sigma$-model of 
\cite{zakharov}---this is a limit of the WZWN 
interaction term, where the integer WZWN coefficient goes to infinity while 
the coupling goes to zero, such that the product of the integer with the cube 
of the coupling is kept constant [\cite{pascos}].  
(N.B.\ Contrast to the interaction of a different model [\cite{plebanski}], 
with derivative structure 0-2-2, which could be regarded as a large-N limit 
of the pseudodual interaction exemplified above by SO(3).)
 
Alternatively, the structure of (\ref{termC}) is linked to what remains of the 
gauge theory instanton density,
\be 
K^0 = \ep^{\mu\nu\kappa}   {\rm Tr} A_\mu \left( 
\partial_{\nu}A_{\kappa} - \frac{1}{4} [A_{\nu},A_{\kappa}]\right) , 
\ee
in the standard space-invariant limit (where the first term vanishes). 

 There is some formal resemblance to membrane interaction terms introduced 
in \cite{zaikov} (in that case a quartic in the $X^\mu$s), which, in turn, 
reflect the symplectic twist of topological terms of \cite{biran} for 
self-dual membranes. But, unlike those interactions, the cubic terms 
considered here do not posit full Lorentz invariance beyond 3-rotational
invariance: they are merely being considered as quantum mechanical systems with
internal symmetry. One may therefore expect this fact to complicate 
supersymmetrization. 

We also succeed in introducing analogous trilinear  interactions for membranes 
embedded in 7-space, which evince similarly interesting properties. 

In what follows, after a brief review of some matrix membrane technology, we
explore the symmetry features of the new terms, and the remarkable symmetry of 
the corresponding vacuum configurations; we describe classical configurations 
of the Nahm type, which we find to be integrable in 3d, as in the conventional 
membrane models. When we interchange the r\^{o}le of dependent and independent 
variables of the 3d PB Nahm equations, we detail how these ``dual" equations 
are solved by Euler's construction, based on harmonic scalar functions.
Our discussion will concentrate on Poisson Brackets, but the 
majority of our results carry over to the Moyal Bracket and matrix commutator 
cases, by dint of the underlying formal analogy. 

However, even though mostly integrable first-order equations are studied 
here, it should be borne in mind that the behaviour of the generic 
solutions to the second-order equations of motion for such systems is often 
chaotic. For example, in the case of QM matrix models (Yang-Mills on a finite 
gauge group, with fields dependent only upon time), characteristic features 
of chaotic behaviour were observed on the solutions of the {\em second-order} 
equations of motion [\cite{savvidy}]. Still, it is not known whether this 
ergodicity persists in the large $N$ limit, i.e.\ the corresponding PB system. 

Subtler topological considerations of special features for
various membranes are not addressed here. Nontrivial boundary terms, 
e.g.~of the type linked to D-branes, are also not considered. 
 
\section{Review of Brackets, Matrix Commutators, and Matrix Membrane Actions } 

Poisson Brackets, Moyal Brackets, and matrix commutators are inter-related 
antisymmetric derivative operations, sharing similar properties, such as 
integration by parts, associativity (hence comportance to the Jacobi identity), 
suitable Leibniz chain rules, etc. They are all representable as commutators 
of associative operators. Much of their technology is reviewed in 
[\cite{moyal};  \cite{ffz};  \cite{ffzjmp}; \cite{hoppe2}]. 

Poisson Brackets act on the ``classical phase-space" of  Fourier-transformed 
color variables, with membrane coordinates $\xi=\alpha,\ \beta$, 
\be
 \{X^\mu,\ X^\nu\}=\frac{\partial X^\mu}{\partial\alpha}\frac{\partial 
 X^\nu}{\partial\beta}-
 \frac{\partial X^\mu}{\partial\beta}\frac{\partial X^\nu}{\partial\alpha} ~.
 \label{PB}
\ee

This may be effectively regarded as the infinitesimal 
canonical transformation on the coordinates $\xi$ of $X^\nu$, generated by 
$\nabla X^\mu\times \nabla $, s.t.\ $(\alpha,\beta) \mapsto (\alpha- 
\partial X_\mu/\partial \beta~,~\beta+\partial X_\mu /\partial \alpha)$,
which preserves the membrane area element $d\alpha d\beta $. This element 
is referred to as a {\it symplectic form} and the class of  transformations 
that leaves it invariant specifies a symplectic geometry; the area preserving
diffeomorphisms are known as {\em SDiff}. 

PBs correspond to $N\rightarrow\infty$ matrix commutators [\cite{hoppe1}].
However, there is a generalization which covers both finite and 
infinite $N$. The virtually unique associative generalization of PBs is the 
Moyal Bracket [\cite{moyal}], 
 \be
 \{ \{ X^\mu,\ X^\nu \} \} =   \frac{1}{\lambda}  \sin\left( \lambda 
\frac{\partial }{\partial\alpha}\frac{\partial }{\partial\beta'}-\lambda 
 \frac{\partial }{\partial\beta}\frac{\partial }{\partial\alpha'} \right) ~
X^\mu (\xi)   X^\nu (\xi')  \Biggl|_{\xi'=\xi}    .
\label{moyal}
 \ee
For $\lambda= 2\pi /N $, \cite{ffz} demonstrate that the Moyal Bracket is 
essentially equivalent to the commutator of SU($N$) matrices---or subalgebras 
of SU($N$), depending on the topology of the corresponding membrane surface
involved in the Fourier-transform of the color indices [\cite{ffzjmp};
\cite{kim}]. 

In the limit $\lambda \rightarrow 0$, the Moyal Bracket goes to the PB, 
i.e.\ $\lambda$ may be thought of as $\hbar$. Thus, when extremely high Fourier
modes are ignored, PBs are seen to represent the infinite $N$ limit. This type
of identification was first noted without benefit of the MB construction by
\cite{hoppe1} on a spherical membrane surface; the foregoing MB limit argument
was first formulated on the torus [\cite{ffz}], but readily extends to other
topologies [\cite{ffz}; \cite{ffzjmp}; \cite{kim}]. 
\begin{figure}
\thicklines 
\unitlength=5mm
\begin{picture}(11,4.2) (-4,0)    
\put(0.7,0){\vector(1,0){7}}
\put(9.5,0.4){\vector(0,1){3.5}}
\put(0.7,0.3){\vector(2,1){7.2}}
\put(0,0){\makebox(0,0)[cc]{MB}} 
\put(5,0.5){\makebox(0,0)[cc]{$\lambda=2\pi/N$}} 
\put(10.7,2){\makebox(0,0)[cc]{$N\rightarrow \infty$}} 
\put(8.8,0){\makebox(0,0)[cc]{SU($N$)}} 
\put(4,3){\makebox(0,0)[cc]{$\lambda\rightarrow 0$}} 
\put(9.5,4.2){\makebox(0,0)[cc]{PB$\sim$SU($\infty$)}} 
\end{picture}
\end{figure}

It would suffice to simply treat these most general Moyal Brackets;  but 
their technical manipulations are sometimes more involved and less familiar, 
so that we also cover the matrix commutator and PB special cases as well, 
redundancy outweighed by pedagogy. 

\cite{floratos3} utilize the abovementioned identification of 
SU($\infty$) with {\em SDiff}  on a 2-sphere, to take the large $N$ limit of 
SU($N$) gauge theory and produce membranes. This procedure was found to be
more transparent on the torus [\cite{ffzjmp}]: the Lie algebra indices
Fourier-conjugate to surface coordinates, and the fields are rescaled Fourier
transforms of the original SU($N$) fields. The group composition rule for them
is given by the PBs and the group trace by surface integration, 
\be
[A_{\mu}, A_{\nu} ] ~\mapsto  \{ a_{\mu},a_{\nu} \} ~;
\ee
\be
F_{\mu \nu}=\partial_{\mu}A_{\nu}-\partial_{\nu}A_{\mu} +[A_{\mu}, A_{\nu}
]~\mapsto 
f_{\mu\nu}(\alpha, \beta) =\partial_{\mu}a_{\nu}-\partial_{\nu}a_{\mu}
+\{ a_{\mu},a_{\nu}\} ~;
\ee
\be
\hbox{Tr} F^{\mu\nu}F_{\mu\nu}\mapsto -{N^3\over 64\pi^4} \int \! d\alpha 
d\beta  ~f^{\mu\nu}(\alpha ,\beta ) ~f_{\mu\nu} (\alpha ,\beta )~.
\ee

But the large $N$ limit need {\em not} really be taken to produce sheet
actions. The Lagrangian with the Moyal Bracket supplanting the Poisson Bracket
is itself a gauge-invariant theory, provided that the gauge transformation also
involves the Moyal instead of the Poisson Bracket: 
\be
\delta a_{\mu} =\partial_{\mu} \Lambda-\{\{\Lambda,a_{\mu}\}\}  ~,
\ee
and hence, by virtue of the Jacobi identity,
\be
\delta f_{\mu\nu}=-\{\{\Lambda,f_{\mu\nu}\}\}      ~.
\ee
Color invariance then follows, 
\be
\delta\int \! d\alpha d\beta ~f^{\mu\nu} f_{\mu\nu}=
-2\int \! d\alpha d\beta ~f^{\mu\nu}\{\{\Lambda,f_{\mu\nu}\}\}=0~.
\ee
The relevant manipulations are specified in \cite{ffzjmp}: the last equality is
evident by integrations by parts, where the surface term is discarded---or
nonexistent if the color membrane surface is closed\footnote{But note this
topological term may be nontrivial for D-membranes.}.  For $\lambda= 2\pi /N $,
this is equivalent to a conventional SU($N$) commutator gauge theory. 

Consider, with \cite{hoppe1}, the SU$(\infty)$ Yang-Mills lagrangian; and 
trivialize all space dependence (through dimensional reduction), leaving 
only time dependence, while preserving all the color-Fourier-space 
(membrane coordinates $\xi=\alpha,~\beta$) dependence of the gauge fields, 
which are now denoted $X^\mu(t,\alpha,\beta)$. Fix the gauge to $X^0=0$,
and consider $\mu,\nu$ to henceforth only range over spacelike values.

The Yang-Mills lagrangian density now reduces to the bosonic membrane 
lagrangian density
\be
 {\l}_{PB} = \frac{1}{2}(\partial_t X^\mu)^2 -\frac{1}{4}\{X^\mu,\ X^\nu\}^2 .
 \label{legba}
\ee
 The PB is also 
the determinant of the tangents to the membrane, so that the conventional 
``potential term" was identified in \cite{ffzjmp} as the Schild-Eguchi string 
lagrangian density [\cite{schild}] (sheet area squared instead of area), 
$ \{ X_{\mu}, X_{\nu}\} \{ X_{\mu}, X_{\nu}\}$. It can be seen that the 
equations of motion of such a string action contain those of Nambu's action. 

Note that, fixing the gauge $X_0=0$ preserves the global color invariance,
i.e.~with a time-independent parameter $\Lambda(\alpha,\beta)$. The action is 
then  invariant under 
\be
 \delta X^\mu = \{\Lambda ,\ X^\mu\} ~.  \label{hatchibombotar}
\ee
By Noether's theorem, this implies the time invariance of the color charge, 
\be
 {\cal Q}_\Lambda   = \int\! d\alpha d\beta ~\Lambda(\alpha,\beta) ~
\{\partial_t X^\mu ,\ X^\mu\} ~.
\ee

The same also works for the Moyal case [\cite{ffzjmp}]. The corresponding Moyal 
Schild-Eguchi term was utilized to yield a ``star-product-membrane'' 
[\cite{hoppe2}],
 \be
 {\l}_{MB}=\frac{1}{2}(\partial_t X^\mu)^2 -\frac{1}{4}\{\{X^\mu,\ X^\nu\}\}^2,
 \label{bondye}
\ee
invariant under 
\be
 \delta X^\mu = \{ \{ \Lambda ,\ X^\mu \} \} ~.
 \label{odoun}
\ee
As argued, this includes the commutator case (QM matrix model), 
\be
 {\l}_{C} = \frac{1}{2}(\partial_t X^\mu)^2 -\frac{1}{4} [X^\mu,\ X^\nu]^2,
 \label{doulamor}
\ee
invariant under 
\be
 \delta X^\mu = [\Lambda ,\ X^\mu ]~.
 \label{varC}
\ee

\section{Cubic Terms for 3 Dimensions}

By suitable integrations by parts, it is straightforward to check that the 
cubic terms (\ref{termPB},\ref{termM},\ref{termC}) in the respective actions, 
$\int dt d\alpha d\beta ~\l$, are 3-rotational invariant, as well as
time-translation invariant and translation symmetric. They are also 
global color invariant, as specified above. 

Now, further consider a plain mass term in the action\footnote{N.B. ~Of the 
type 
that may arise as a remnant of space gradients in compactified dimensions.}, 
\be
 {\l}_{3dPB} = \frac{1}{2}(\partial_t X^\mu)^2 -\frac{1}{4}\{X^\mu,\ X^\nu\}^2
 -\frac{m}{2} \ep^{\mu\nu\kappa} X^\mu \{ X^\nu ,\ X^\kappa \} -
 \frac{m^2}{2}(X^\mu)^2  .
 \label{bokor}
\ee
The second order equation of motion,
\be
 \partial_t^2 X^\mu  = - m^2 X^\mu -\frac{3m}{2}\ep^{\mu\nu\kappa} 
 \{ X^\nu ,\ X^\kappa \}-  \{  X^\nu  ,  \{ X^\mu ,\ X^\nu \}\}~ ,
 \label{bossou}
\ee
follows not only from extremizing the action, but also results from a 
first-order equation of the Nahm (self-dual) type [\cite{neq}], 
albeit complex, 
\be
 \partial_t X^\mu = imX^\mu +\frac{i}{2}\ep^{\mu\nu\kappa} \{ X^\nu ,\ 
 X^\kappa \}~.
 \label{nahm}
\ee
These equations hold for PBs, as well as for MBs and matrix commutators. 

For solutions of this first-order  equation, the conserved energy vanishes.
In general, however, such solutions are not real, and do not provide 
absolute minima for the action---the reader may consider the simple 
harmonic oscillator to illustrate the point. Nonetheless, the lagrangian 
density can be expressed as a sum of evocative squares with positive relative 
signs, since the potential in (\ref{bokor}) is such a sum, 
\be
 {\l}_{3dPB} = \frac{1}{2}(\partial_t X^\mu)^2 -\frac{1}{2}\left( m X^\mu 
 +\frac{1}{2}\ep^{\mu\nu\kappa} \{ X^\nu ,\ X^\kappa \}\right) ^2.\label{mambo} 
\ee
By integration by parts in the action $\int dt d\alpha d\beta ~ {\l}_{3dPB} $,
the lagrangian density itself can then be altered to 
\be
 {\l}_{3dPB}\cong - \frac{1}{2}\left( i\partial_t X^\mu +m X^\mu +\frac{1}{2}
 \ep^{\mu\nu\kappa} \{ X^\nu ,\ X^\kappa \}\right) ^2 ,
\ee 
just like the conventional bosonic membrane lagrangian density\footnote{The 
congruence symbol, $\cong$, denotes equivalence up to surface terms, which, 
e.g., vanish for a closed surface; again, consideration of D-membranes would 
proceed separately.}. Naturally, the complex-conjugate versions of the above 
are equally valid.      

\section{Vacuum Configurations and their Symmetry} 

The minimum of the conventional matrix membrane trough potential favors 
alignment of the dynamical variables $X^\mu$s. 
The mass parameter introduced above parameterizes a partial trough symmetry 
breaking\footnote{\cite{curtmcc} have considered a system equivalent to the 
limit of constant $X^3=$ in this model.}, but does not lift 
``dilation" invariance, seen as follows. 

The static ($t$-independent) minima for the action (vacuum configurations) 
are solutions of
\begin{equation}
 mX^{\mu }+\frac{1}{2}\epsilon ^{\mu \nu \kappa }\{X^{\nu },\ X^{\kappa}\}
 =0\;.  \label{static}
\end{equation}
The previously considered case, $m=0$, is easily solved by ``color-parallel"
configurations. But for $m\neq 0$, the static solutions must lie on a 
2-sphere, since from the previous equation
\begin{equation}
 X^{\mu }\frac{\partial X^{\mu }}{\partial \alpha }=0=X^{\mu }\frac{\partial
 X^{\mu }}{\partial \beta }\;,
\end{equation}
so 
\be
X^{\mu }X^{\mu }=R^2~,
\ee
an unspecified  constant\footnote{Hence 
$\epsilon^{\mu \nu \kappa } X^\mu \{ X^{\nu },\ X^{\kappa} \}=-2mR^2$.}.
However, from (\ref{static}), note that both $m$ and 
also $R$, the scale of the $X^\mu$s, can
be absorbed in the membrane coordinates $\xi$ and will not be specified by
the solution of (\ref{static}).  

Indeed, solving for one coordinate component on this sphere, say 
\be
X(Y,Z)=\pm  \sqrt{R^{2}-Y^{2}-Z^{2}}~,
\ee
reduces the three equations (\ref{static}) to one. Namely, 
\begin{equation}
 \{Z,Y\}=m\sqrt{R^{2}-Y^{2}-Z^{2}}~,
\end{equation}
on the positive $X$ branch ($m\mapsto -m$ on the negative $X$ branch). This
last equation is solved by 
\be
 Z=\alpha ~, \qquad Y=\sqrt{R^{2}-\alpha ^{2}}\sin (m\beta )~.
\ee
\begin{figure}
\thicklines 
\unitlength=4mm
\begin{picture}(11,10.5) (-3,0)    
\put(0,0){\line(0,1){10}}
\put(0,10.5){\makebox(0,0)[cc]{$Z$}}
\put(0,0){\line(1,0){11}}
\put(11.5,0){\makebox(0,0)[cc]{$X$}}
\put(0,0){\line(5,2){11}}
\put(11,3.7){\makebox(0,0)[cc]{$Y$}}
\put(0,0){\vector(1,2){4.6}}
\put(0,0){\vector(4,1){4.5}}
\put(0,9){\makebox(0,0)[cc]{o}} 
\put(1.9,0){\makebox(0,0)[cc]{o}} 
\put(3,1.2){\makebox(0,0)[cc]{o}} 
\put(-1,9){\makebox(0,0)[cc]{$\alpha$}}
\put(4,9){\makebox(0,0)[cc]{$R$}}
\put(3,0.3){\makebox(0,0)[cc]{$m\beta ~)$}}
\end{picture}
\end{figure}
One can then interpret $m\beta $ as the usual azimuthal angle
around the $Z$-axis. Hence, $-\pi /2\leq m\beta \leq \pi /2$ and $-R\leq
\alpha \leq R$ covers the $X\geq 0$ hemisphere completely. The other
hemisphere is covered completely by the negative $X$ branch. Since $R$ is
not fixed, it amounts to an unlifted residual trough dilation degeneracy.

All static solutions are connected to this explicit one by
rescaling $R$ and exploiting the equation's area-preserving diffeomorphism
invariance for $\xi =(\alpha ,\beta )$.

\section{Nahm's Equation and its Lax Pairs} 

The first order equation, (\ref{nahm}), simplifies upon changing variables to 
\be
\tau ={\E^{imt}\over m} ~, \qquad\qquad
X^\mu =\E^{imt} Y^\mu ,
\ee 
and reduces to the conventional PB version [\cite{ward},\cite{leontaris}] 
of Nahm's equation [\cite{neq}],  
\be
 \partial_\tau Y^\mu  = \frac{1}{2}\ep^{\mu\nu\kappa} \{ Y^\nu ,\ Y^\kappa \}~.
 \label{orignahm}
\ee
This one does have real solutions, and can be linearized by Ward's 
transformation [\cite{ward}]. However, the action (\ref{mambo}) does 
not reduce to the conventional one upon these
transformations\footnote{Likewise, the second order equations of motion only
reduce to $ \partial_{\tau}^2  Y^\mu  =  \{ Y^\nu, \{ Y^\mu , Y^\nu \} \}+  
\frac{3}{\tau}( \partial_\tau Y^\mu  - \frac{\ep^{\mu\nu\kappa}} {2}  
\{ Y^\nu , Y^\kappa \})$.}.  

Moreover, note 
\be
\partial_\tau Y^\mu \partial_\tau Y^\mu = { \partial (Y^1 ,Y^2, Y^3)\over 
\partial (\tau ,\alpha,\beta)} ~,
\ee
\be
\partial_\tau Y^\mu \partial_\xi Y^\mu =0 ~.
\ee
 
One may further utilize the cube root of unity, $\omega =\exp (2\pi i/3)$,
to recast\footnote{N.B.~$\omega (\omega -1)$ is pure imaginary.} 
(\ref{orignahm}),
\be
 L\equiv \omega  Y^1 + \omega ^2 Y^2 + Y^3~, \quad 
 \overline{L}=\omega^2  Y^1 + \omega Y^2 + Y^3~, \quad 
 M\equiv Y^1 + Y^2 + Y^3~, 
\ee
\be
 \omega (\omega -1)~ \partial_\tau L= \{  M~,~ L\}~,\qquad 
 \omega (\omega -1)~ \partial_\tau \overline{L}= 
 -\{  M~,~ \overline{L} \}~, \label{erzulie}
\ee
\be 
 \omega (\omega -1)~ \partial_\tau M= \{ L ~,  \overline{L} \}~,  
\label{ezulie}
\ee
which thus yields an infinite number of complex time-invariants,
\be 
Q_n=\int\! d\alpha d\beta ~L^n   ~,  
\ee
for arbitrary integer power $n$, as the time derivative of the integrand 
is a surface term. (This is in complete analogy with the standard case of 
commutators.)  The link to classical integrability is discussed next.

Equations (\ref{erzulie},\ref{ezulie}) amount to one complex and one real 
equation, but these are known to be further capable of compacting into just one
by virtue of an arbitrary real spectral parameter $\zeta$, e.g. as introduced
in \cite{leontaris}: 
\be
 H\equiv {i\over \sqrt{2} \omega(\omega-1)}
 \left( \zeta L - {\overline{L} \over \zeta}\right) ~, \qquad \qquad 
 K\equiv i \sqrt{2} M + \zeta L + {\overline{L} \over \zeta} ~, 
\ee
\be
 \partial_\tau K= \{ H, K \}~.
\ee
Likewise, this Lax pair\footnote{Note the wave solutions 
$H=\alpha,~ K= f(\beta + \tau)$.},
analogous to \cite{hitchin}, leads to a one-parameter family of 
time-invariants\footnote{ 
L. Dickey calls our attention to the parameter introduced in the 
Lax pair for the generalized Euler equations by  S. Manakov, Funct Anal Appl 
{\bf 10} (1976) 328, which parallels that in Hitchin's Lax pair for the 
Nahm equations.},
\be
{\cal Q}_n (\zeta) =\int\! d\alpha d\beta ~ K^n ~,  
\ee
conserved for all $n$ and $\zeta$. 

It also yields the usual Lax isospectral flow. PBs (and, {\em mutatis 
mutandis}, MBs) can be recast into commutators of suitable associative 
operators 
\be
{\cal K}\equiv \nabla K \times \nabla ,\quad \qquad 
{\cal H}\equiv \nabla H \times \nabla ,\quad \qquad 
 \nabla \equiv \left( {\partial \over \partial\alpha}, 
{\partial\over \partial\beta}\right) ~, 
\ee
such that:
\be 
\partial_\tau  {\cal K}  = {\cal H} {\cal K} -{\cal K} {\cal H} ~.\label{vodoun}
\ee
As a consequence, the spectrum of ${\cal K}$ is preserved upon time evolution 
by the (pure imaginary) ${\cal H}$:
\be
\partial_\tau \psi= {\cal H} \psi ~,
\ee
since time-differentiating 
\be
{\cal K} \psi= \lambda\psi 
\ee
and applying (\ref{vodoun}) yields 
\be  
(\partial_\tau \lambda) ~\psi = (\partial_\tau  {\cal K}) \psi+ 
{\cal K}\partial_\tau  \psi- \lambda~\partial_\tau \psi =0  ~.
\ee 
This isospectral flow then provides integrability for (\ref{nahm}), as in the 
case of the matrix commutator Nahm equation.

The discussion so far also carries over to plain matrix commutators or Moyal 
Brackets as well, with suitable adaptations for the associative operators 
involved, e.g. ~a $\star$-product structure [\cite{ffz}; \cite{hoppe2}], 
\be
\bbbk\equiv {1\over 2i\lambda}~ \E^{i\lambda \nabla ' \times\nabla} ~ K(\xi') 
\Biggl|_{\xi'=\xi} .
\ee

For the rest of this section, we shall restrict our attention to the PB case 
only. \cite{ward} has solved the PB Nahm equation (\ref{orignahm}) implicitly 
through twistor linearization. Another solution procedure for (\ref{orignahm})
may be found by interchanging the r\^oles of dependent and independent
variables: the equations then take the dual form 
\be
\frac{\partial \tau}{\partial Y^1}=\frac{\partial \alpha}{\partial Y^2}
\frac{\partial \beta}{\partial Y^3}-
\frac{\partial \alpha}{\partial Y^3}\frac{\partial \beta}{\partial Y^2} ~,
\label{replace}
\ee
together with cyclic permutations, i.e. 
\be
\partial_\mu \tau =\ep^{\mu\nu\rho} \partial_\nu \alpha ~\partial_\rho \beta ~.
\label{philophage}
\ee

Cross-differentiation produces integrability conditions 
\be
\partial_\mu ( \partial_\kappa \alpha ~\partial_\mu \beta - 
\partial_\mu \alpha ~\partial_\kappa \beta)    =0~. 
\label{podular}
\ee
Another evident consistency condition is  
\be
\partial^2 \tau = 0~.
\label{houngan}
\ee

In fact, any harmonic function $\tau(Y^1, Y^2,Y^3)$, readily yields 
$\alpha,\ \beta$. Euler appreciated (1770) that any continuous differentiable 
divergenceless vector field may be represented locally as a cross product of 
two gradients [\cite{ericksen}]. The problem of solving the inverse 
Nahm equation (\ref{philophage}) then reduces to Euler's construction, 
given an arbitrary harmonic function $\tau$. 

To determine scalar fields $\alpha$ and $\beta$, note that 
\be
\nabla \alpha \cdot \nabla  \tau =0= 
\nabla \beta \cdot \nabla  \tau ~,
\ee
so that $\nabla  \tau$ lies on surfaces of constant $\alpha $ and $\beta   $.
One may then choose the surface of constant $\alpha $ through $\nabla  \tau$
arbitrarily, and further choose an arbitrary continuous vector field ${\bf V}$
s.t.\ ${\bf V}\cdot \nabla \alpha\neq 0$. One may thus integrate on 
the constant-$\alpha $ surface to obtain $\beta$:
\be
\beta= \int d{\bf Y} ~\cdot  {  \nabla \tau \times {\bf V} \over {\bf V}\cdot 
\nabla \alpha  } ~  . 
\ee

Evidently, solutions of these dual equations 
\be
\partial_\mu f =\ep^{\mu\nu\rho} \partial_\nu  g  ~ \partial_\rho h ~,
\label{philopote}
\ee
for $f(Y^1,Y^2,Y^3)$ produce constants of the motion $\int d^2 \xi~ f$,
beyond those already found by the Lax procedure, for 
the {\em original} equation (\ref{orignahm}), in illustration of a phenomenon 
noted in \cite{strach}, as it is straightforward to verify that 
\be
{\D f\over \D \tau}=\{g,h\}~.
\ee
From the foregoing discussion, $f$ need only solve Laplace's equation 
(\ref{houngan}): any harmonic function $f(Y^1,Y^2,Y^3)$ yields a conserved 
density for (\ref{orignahm}) by also satisfying (\ref{philopote}).
To belabor the point, by virtue of Helmholtz's theorem, a divergenceless 
3-vector $\nabla f$ is representable as a curl of another vector ${\bf A}$. 
On the other hand, an arbitrary 3-vector can also be represented in terms of
three scalars (``Monge potentials") by means of the non-unique 
``Clebsch decomposition" of that vector as ${\bf A}=g\nabla h + \nabla u $, 
in a trivial extension of Euler's construction just outlined [\cite{ericksen}]. 

\section{Membrane Embedding in 7 Dimensions}

Remarkably, the same type of cubic interaction term may also be introduced 
for a membrane embedded in 7 space dimensions. An antisymmetric, 
self-dual 4-tensor in 8 
dimensions, $f_{\mu\nu\rho\sigma}$ was invoked by \cite{cor} as an
8-dimensional analog of the 4-dimensional fully antisymmetric tensor
$\epsilon_{\mu\nu\rho\sigma}$. Some useful technology for the manipulation
of this tensor (which has 35 nonzero components, is linked to Cayley's 
octonionic structure constants, and is invariant under a 
particular $SO(7)$ subgroup of $SO(8)$) can be found in \cite{gursey}; 
in particular, the identity
\be
 f^{0\mu\nu\kappa} f^{0\mu\lambda\rho}= f^{\nu\kappa\lambda\rho} 
 +\delta^{\nu\lambda} \delta^{\kappa\rho} -\delta^{\kappa\lambda}
  \delta^{\nu\rho}~.\label{gurseyidentity} 
\ee

By analogy with (\ref{orignahm}), we postulate a first-order 
equation\footnote{At present, we are not in a position to comment on its 
integrability or lack thereof.},
\be
 \partial_\tau Y^\mu -{f^{0\mu\nu\kappa}\over 2}  \{ Y^\nu ,\ Y^\kappa \} =0~. 
 \label{orignahm2}
\ee
The indices run from $\mu =1$ to $\mu=7$, since we are working in a gauge 
where $Y^0=0$.
The second order equation arising from iteration of (\ref{orignahm2}), by 
virtue of the above identity, as well as the Jacobi identity, is 
\be
 \partial_{\tau\tau} Y^\mu  = -\{\{ Y^\mu ,\ Y^\nu \}  ,\ Y^\nu \}  ~.
 \label{baronsamedi}
\ee

This arises from the lagrangian density 
\be
 {\l}_{7dPB}=\frac{1}{2} \left( \partial_\tau Y^\mu\right)^2
 +\frac{1}{4}\{Y^\mu,\   Y^\nu\}^2 ~ . \label{tapitor}
\ee
As in the 3-dimensional case, this action reduces to a sum of squares 
with positive relative signs, up to a mere surface term,
\be
 \frac{1}{2}\! \left( \partial_\tau Y^\mu  -  {f^{0\mu\nu\kappa}\over 2}  
 \{ Y^\nu ,\ Y^\kappa \}\right)^2  \! 
 ={\l}_{7dPB} - f^{0\mu\nu\kappa} \partial_\tau Y^\mu\partial_\alpha 
 Y^\nu\partial_\beta Y^\kappa \cong {\l}_{7dPB}. \label{top}
\ee
In this lagrangian density, apparent extra terms 
$f^{\mu\nu\rho\kappa}  \{ Y^\mu ,\ Y^\nu\} \{ Y^\rho ,\ Y^\kappa \}$ 
induced by the identity (\ref{gurseyidentity}) have, in fact, 
vanished, by virtue of the identity,
\be
 \{f,\ g\}\{h,\ k\} + \{f,\ h\}\{k,\ g\}+\{f,\ k\}\{g,\ h\}\equiv 0~,
 \label{saturday}
\ee
which holds for Poisson Brackets on a 2-dimensional phase-space---but not for
matrix commutators nor Moyal Brackets\footnote{ D. Fairlie 
and A. Sudbery (1988, unpublished). It follows from 
$\epsilon^{[jk} \delta^{l]m}=0$, whence  
$\epsilon^{[jk} \epsilon^{l]m}=0$ for these membrane symplectic coordinates.}.
This cancellation works at the level of the lagrangian density for the PB case.

However, note that even for ordinary matrices the corresponding term would 
vanish in the traced lagrangian, by the cyclicity of the trace pitted against 
full antisymmetry, 
\be
f^{\mu\nu\kappa\rho }\hbox{Tr} X^\mu X^\nu X^\kappa X^\rho =0~.
\ee
Likewise, the corresponding interaction for Moyal Brackets,
\be
f^{\mu\nu\kappa\rho }\int\! d^2 \xi ~\{\{X^\mu, X^\nu\}\}
\{\{ X^\kappa, X^\rho \}\}~ ,
\ee
is forced by associativity to reduce to a surface term, 
vanishing unless there are contributions from surface boundaries or
D-membrane topological numbers involved. (Shortcuts for the manipulation of 
such expressions underlain by $\star$-products can be found, e.g.,  in 
\cite{hoppe2}.) The cross terms involving time derivatives are 
expressible as divergences, as in the 3-dimensional case, and hence may give 
rise to topological contributions.

As a result, (\ref{orignahm2}) is the Bogomol'nyi minimum of the action of 
(\ref{tapitor}) with the bottomless potential.

As in the case of 3-space, the conventional membrane signs can now be 
considered (for energy bounded below), and a symmetry breaking 
term $m$ introduced, to yield 
\be
{\l}_{7dPB}\cong - \frac{1}{2}\left(- i\partial_t X^\mu +m X^\mu +
{f^{0\mu\nu\kappa}\over 2}  \{ X^\nu ,\ X^\kappa \}\right) ^2.
\ee 

This model likewise has 7-space rotational invariance, and its vacuum 
configurations are, correspondingly, 2-surfaces lying on the spatial 6-sphere 
embedded in 7-space: $X^{\mu }X^{\mu }=R^{2}$. But, in addition, because 
of (\ref{gurseyidentity}), these surfaces on the sphere also satisfy the
trilinear constraint 
\be
 f^{\lambda \mu \nu \kappa }X^{\mu }\{X^{\nu },\
 X^{\kappa }\}=0 \enspace ,
\ee
for $\lambda\neq 0$.   (For $\lambda=0$ this trilinear is $-2mR^2$.) 
 
\end{document}